\documentclass[11pt]{article}
\usepackage[T1]{fontenc}

\usepackage[utf8]{inputenc}
\usepackage[english]{babel}
\usepackage[margin=2.3cm]{geometry} %
\usepackage{graphicx} %
\usepackage{amsthm, amsmath, amssymb} 
\usepackage{tcolorbox} 
\usepackage{placeins} %
\usepackage[numbers,sort&compress]{natbib} %
\usepackage{lineno}	
\usepackage{appendix} %

\usepackage{authblk} 
 \usepackage{setspace} %
\usepackage{pifont}

\usepackage{hyperref} %
\newcommand{\papertitle}{Elasticity of bidisperse attractive particle systems}

\newcommand{\ie}{{\textit{i.e.}}}
\newcommand{\eg}{{\textit{e.g.}}}

\newcommand{\shearmodulus}{G}

\newcommand{\coordinationnumber}{\langle Z\rangle}
\newcommand{\approxz}{\tilde{Z}}

\newcommand{\labelL}{_\mathrm{L}}
\newcommand{\labelS}{_\mathrm{S}}
\newcommand{\labelLL}{_\mathrm{LL}}
\newcommand{\labelSL}{_\mathrm{SL}}
\newcommand{\labelSS}{_\mathrm{SS}}

\newcommand{\labelmono}{_\mathrm{mono}}

\newcommand{\packingfraction}{\phi}
\newcommand{\packingfractionL}{\packingfraction\labelL}
\newcommand{\packingfractionS}{\packingfraction\labelS}

\newcommand{\ljpotential}{U}
\newcommand{\scalingparameter}{\alpha}
\newcommand{\ljshift}{\Delta}
\newcommand{\ljsigma}{\sigma_\mathrm{lj}}
\newcommand{\ljheq}{h_\mathrm{eq}}
\newcommand{\ljeps}{\epsilon_\mathrm{lj}}
\newcommand{\ljstiffness}{k}
\newcommand{\ljstiffnessLL}{\ljstiffness\labelLL}
\newcommand{\ljstiffnessSS}{\ljstiffness\labelSS}
\newcommand{\ljstiffnessSL}{\ljstiffness\labelSL}

\newcommand{\radius}{R}
\newcommand{\radiusL}{\radius\labelL}
\newcommand{\radiusS}{\radius\labelS}
\newcommand{\normalizedsmallradius}{\overline{\radius}}

\newcommand{\boxsize}{L}
\newcommand{\volume}{V}
\newcommand{\boxvolume}{\volume_\mathrm{box}}
\newcommand{\particlevolume}{\volume_\mathrm{p}}
\newcommand{\volumeL}{\volume\labelL}
\newcommand{\volumeS}{\volume\labelS}
\newcommand{\smallparticlevolumefraction}{\overline{\volume}\labelS}

\newcommand{\particlenumber}{N}
\newcommand{\particlenumberS}{\particlenumber\labelS}
\newcommand{\particlenumberL}{\particlenumber\labelL}

\newcommand{\connectionumber}{N}
\newcommand{\connectionnumberLL}{\connectionumber\labelLL}

\newcommand{\connectionnumberSL}{\connectionumber\labelSL}

 \graphicspath{ {./} }
\begin{document}

\title{\papertitle}
\author[1]{Yaqi Zhao}
\author[1]{Antoine Sanner} %
\author[1]{Luca Michel}  %

\author[1]{David S. Kammer\thanks{Corresponding Author: dkammer@ethz.ch}}

\affil[1]{Institute for Building Materials, ETH Zurich, Switzerland}

\maketitle

\section*{Abstract}
Bidisperse particle systems are common in both natural and engineered materials, and it is known to influence packing, flow, and stability. However, their direct effect on elastic properties, particularly in systems with attractive interactions, remains poorly understood. Gaining insight into this relationship is important for designing soft particle-based materials with desired mechanical response. In this work, we study how particle size ratio and composition affect the shear modulus of attractive particle systems. Using coarse-grained molecular simulations, we analyze systems composed of two particle sizes at fixed total packing fraction and find that the shear modulus increases systematically with bidispersity. To explain this behavior, we develop two asymptotic models following limiting cases: one where a percolated network of large particles is stiffened by small particles, and another where a small-particle network is modified by embedded large particles. Both models yield closed-form expressions that capture the qualitative trends observed in simulations, including the dependence of shear modulus on size ratio and relative volume fraction. Our results demonstrate that bidispersity can enhance elastic stiffness through microstructural effects, independently of overall density, offering a simple strategy to design particle-based materials with tunable mechanical properties.

\vspace{5mm} \noindent \textbf{Keywords:}
particle systems; bidispersity; elastic properties; 

\newpage
\noindent \textbf{Highlights:}
\begin{itemize}
    \item Bidispersity increases shear modulus even at fixed total packing fraction
    \item Simulations reveal a strong dependence on size ratio and composition
    \item Two asymptotic regimes capture limiting behaviors of bidisperse systems
    \item Closed-form expressions predict stiffening from microstructural mechanisms
\end{itemize}

\newpage
\section{Introduction} \label{sec:introduction}

Polydisperse particle systems are ubiquitous in both natural and industrial contexts, appearing in materials such as soils, powders, suspensions, and a wide range of products in the cosmetics, food, and construction industries~\cite{ren2018grain,hlobil2022surface,afoakwa2008effects,guerra2023influence,cassayre2024optimization,hsu2021exploring,muller2025tuning}. These systems consist of particles that vary in size and shape, and their macroscopic mechanical behavior emerges from a complex interplay of diverse factors including particle shape and size, packing configuration, connectivity, and interparticle forces~\cite{PhysRevE.60.4551, PhysRevE.76.021301, herrmann2003searching, newhall2011statistical, zhang2015structure, shaebani2012influence, wackenhut2005shearing, PhysRevLett.102.178001, cantor2018rheology, goncu2010constitutive, ioannidou2016mesoscale, ioannidou2016crucial,zaccone2009elasticity,whitaker2019colloidal}. These factors give rise to intricate spatial arrangements and force networks that govern a material's mechanical response~\cite{torquato2002random, ioannidou2016crucial}. Despite the prevalence and importance of polydisperse particle systems, the fundamental mechanisms governing their mechanical behavior remain poorly understood, both for size and shape polydispersity. 

While many practical systems feature variability in both properties, even the simplified case of spherical particles with broad size distributions presents significant challenges. To address this, bidisperse systems offer a simplified model framework that retains essential features of polydispersity while reducing its complexity. By allowing controlled variation of particle size while maintaining fixed shape, bidisperse systems serve as a canonical model for understanding the effect of size variability on mechanical behavior.

To understand the role of bidispersity, it is important to distinguish between repulsive and attractive particle systems. In repulsive systems such as dry granular materials, the mechanical behavior is governed by contact forces and friction, and the addition of smaller particles is known to improve packing efficiency, enhance connectivity, and increase overall stability~\cite{mascioli_defect_2017, chaki_theoretical_2024, kumar_tuning_2016,yuan_connecting_2021}. These effects are generally attributed to a densification mechanism, where smaller particles fill the interstitial space between larger ones. In contrast, attractive systems such as colloidal gels or suspensions are governed by short-range cohesive forces rather than frictional contacts, leading to fundamentally different mechanisms of force transmission and structural organization. As a result, the insights gained from repulsive systems may not directly translate, and the influence of bidispersity on the mechanical behavior of attractive particle systems remains insufficiently understood.

Bidispersity has been shown to affect the phase behavior, rheology, and structural organization in attractive spherical systems, such as colloidal materials~\cite{yang_tunable_2013, jiang_flow-switched_2022, schulte_microgels_2022}. Introducing a second particle size can alter packing efficiency, promote the formation of particle bridges, or generate depletion effects that strongly influence stress transmission and mechanical stability~\cite{xu_jamming_2023, cao_solidliquid_2015, ji_interaction_2013, manoharan_colloidal_2015, jie_chen_xuewu_wang_steven_r_kline_and_yun_liu_size_nodate,orefice2023numerical}. These studies demonstrate that bidispersity can play a central role in organizing colloidal structures and shaping their mechanical response.

Most of the reported effects, however, concern the fluid-solid transition of particle systems~\cite{yang_tunable_2013,xu_jamming_2023,jiang_flow-switched_2022}. In contrast, much less is known about how bidispersity influences the elastic response of a solidified attractive particle systems. It has been speculated that the addition of smaller particles could lead to stiffening~\cite{yang_tunable_2013,schulte_microgels_2022}, but available studies often involve simultaneous changes in packing fraction, network connectivity, or phase structure, making it difficult to isolate the specific role of bidispersity.

In this paper, we show that bidisperse attractive spherical packings exhibit a significant stiffening effect compared to monodisperse systems at the same total packing fraction, indicating that the increase in shear modulus cannot be attributed solely to densification. Using coarse-grained molecular simulations, we systematically examine how the shear modulus depends on particle size ratio and composition in systems with short-range attractive interactions. Our analysis reveals that bidispersity enhances elasticity through two distinct mechanisms, each associated with a limit of the system: one where a small number of fine particles locally stiffens a network of larger ones, and another where a sparse population of large particles reinforces a network formed by smaller ones. These results provide a unified framework for understanding how bidispersity controls elasticity in attractive particle systems, independent of overall density.

The remainder of this paper is structured as follows. In Section~\ref{sec:mono}, we present the monodisperse case as a baseline for comparison. Section~\ref{sec:bi} details the results from bidisperse systems, highlighting the observed stiffening mechanisms, which are then discussed in Section~\ref{sec:discussion}.

\section{Monodisperse particle system} \label{sec:mono}

We begin by analyzing monodisperse systems as a reference case. Specifically, we examine how the shear modulus varies with particle radius and packing fraction, providing a baseline for comparison with bidisperse systems modeled and analyzed in Sec.~\ref{sec:bi}.

\subsection{Problem statement \& method} \label{sec:mono:problem_statement_method}

We consider a dense monodisperse system of spherical particles with short-ranged attractive interactions and evaluate the tangent elastic modulus. The model consists of a three-dimensional representative volume element (RVE) that is a box of side length $\boxsize = 100$ (all quantities are unit-less) and volume $\boxvolume = \boxsize^3$. The box contains a variable number $\particlenumber$ of particles with radius $\radius$, which results in a packing fraction of $\packingfraction = \particlevolume / \boxvolume$ with the total volume of particles being $\particlevolume = 4 \pi \particlenumber \radius^3 / 3$. 

The particle interaction is modeled by a shifted Lennard-Jones (LJ) potential
\begin{equation}
	\ljpotential(r) = 4\ljeps\left[\left(\frac{\ljsigma}{r-\ljshift}\right)^{12}-\left(\frac{\ljsigma}{r-\ljshift}\right)^6\right],
    \label{eq:LJ}
\end{equation}
where $r$ is the center-to-center particle distance, $\ljeps$ is the depth of the potential well, $\Delta$ is the shift of the equilibrium position, and $\ljsigma$ is the distance where the shifted potential is zero. Here, we choose $\ljsigma = \ljheq / 2^{1/6}$ with $\ljheq = \scalingparameter /2$ to model short-ranged interactions and $\scalingparameter = \radius / 2$ to ensure short-rangedness for varying particle sizes. Further, we set $\ljshift = 2\radius - \ljheq$ to account for the finite size of the particles. This defines the equilibrium position between two particles at $2\radius$, \ie, when the particles are touching each other. Finally, we set $\ljeps = \scalingparameter^2 \radius$, such that the interaction stiffness between two particles at equilibrium is proportional to their radius:
\begin{equation}
    \ljstiffness = \left. \frac{\partial^2 \ljpotential}{\partial r^2} \right|_{r = 2\radius} \propto \radius ~.
    \label{eq:ktoR}
\end{equation}

We initialize the model by randomly placing $N$ particles into the RVE. Then, we use an artificial soft relaxation process~\cite{ness2023simulating} to eliminate overlapping of particles. This was performed using a constant NVE integration with a soft repulsive granular model using LAMMPS~\cite{LAMMPS} (detailed parameters provided in Appendix~\ref{appendix:solver}). While this preparatory process is not strictly necessary, it enhances convergence and efficiency of the subsequent equilibrium calculation.

The equilibrium state of the RVE is found using the Fast Inertial Relaxation Engine (FIRE) algorithm, chosen for its robustness and efficiency in navigating complex energy landscapes. The resulting configuration represents a stable, load-bearing microstructure (detailed setup provided in Appendix~\ref{appendix:solver}).

We compute the elastic properties of the equilibrated monodisperse particle systems (see Fig.~\ref{fig:mono}a) by considering the collective contribution of all pairwise interactions. The shear modulus is determined from the Born tensor $C^B_{i,j}$, given by~\cite{theodorou1986atomistic,lutsko1989generalized}
\begin{equation}
    C^B_{i,j} = \frac{1}{\boxvolume} \frac{\partial^2U}{\partial\varepsilon_i\partial\varepsilon_j} ~,
\end{equation}
where $\varepsilon_i$ and $\varepsilon_j$ are the components of the strain tensor $\boldsymbol{\varepsilon}$. To compute $\partial^2U / \partial\varepsilon_i\partial\varepsilon_j$, we use the chain rule to introduce $\partial^2 U / \partial r^2$ and $\partial r / \partial \varepsilon_i$, which are obtained analytically from the particle positions in the equilibrated state without applying any deformation on the system (details provided in Appendix~\ref{appendix:modulus_computation}). We directly estimate the stiffness tensor $\mathbf{C} = \langle \mathbf{C}^B\rangle$, where the angle brackets represent the average over all realizations. For an isotropic system, the shear modulus $G$ is directly obtained from the shear-related components of $\mathbf{C}$: $G = C_{44}=C_{55}=C_{66}$ (given Voigt notation). Since our simulation domain is not perfectly isotropic due to its finite size, we compute the shear modulus $\shearmodulus$ as their average, given by $G = (C_{44}+C_{55}+C_{66})/3$, to mitigate the anisotropy induced by heterogeneity. This approach directly links the microscopic interparticle potentials to the macroscopic elastic response of the equilibrated particle system.

\subsection{Baseline results: elasticity of monodisperse particle systems} \label{sec:mono:results}

A monodisperse system of spherical particles is, for a given interaction, defined by two parameters: the particle radius $\radius$ and the packing fraction $\packingfraction$. Here, we analyze the effect of these parameters on the elasticity of such systems.

\begin{figure}
    \centering
    \includegraphics[width=1\linewidth]{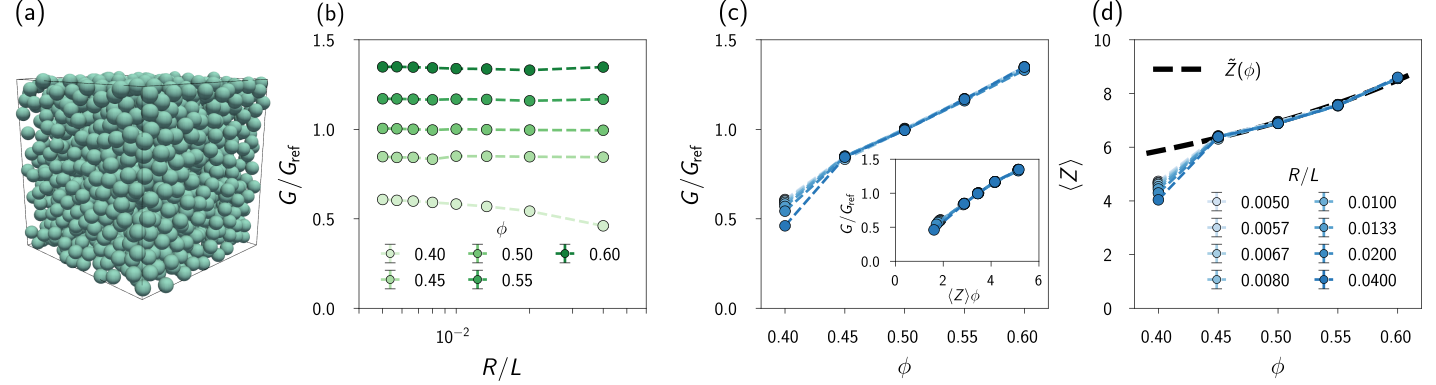}
    \caption{The shear modulus of monodisperse systems with spherical particles is independent of the particle radius and solely determined by the packing fraction. \textbf{(a)}~3D visualization of an equilibrated monodisperse system at a packing fraction $\packingfraction = 0.5$ and a particle radius $\radius/\boxsize=0.02$. \textbf{(b)}~Shear modulus \(\shearmodulus\), normalized by the reference shear modulus \(\shearmodulus_\mathrm{ref}\) (calculated as the average shear modulus at $\packingfraction = 0.5$ and $\radius/\boxsize = 0.005$), plotted as a function of the particle radius \(\radius\) for various packing fractions \(\packingfraction\). Reported values are the average over 10 realizations and error bars represent the 95\% confidence interval. Error bars are smaller than the marker size. \textbf{(c)}~Normalized shear modulus as a function of $\packingfraction$ for different $\radius$. Legend is given in panel d. (Inset) Normalized shear modulus plotted against the product \(\coordinationnumber\packingfraction\) \textbf{(d)}~Average coordination number \(\coordinationnumber\) as a function of packing fraction for various radii. The black dashed line represents a fit using Eq.~\ref{eq:fit}.
    }
    \label{fig:mono}
\end{figure}

First, we consider a fixed packing fraction and vary the particle radius. The results show that the shear modulus \(\shearmodulus\) remains essentially independent of \(\radius\) (see \eg, $\packingfraction = 0.5$ in Fig.~\ref{fig:mono}b). This observation holds for all packing fractions $\packingfraction > 0.4$. At lower values of $\packingfraction$, particularly at $\packingfraction=0.4$, we observe a slight decrease of $\shearmodulus$ for larger $\radius$ (see Fig.~\ref{fig:mono}b). This decrease is likely due to the characteristic cluster size becoming comparable to the finite size of the simulation box for packing fractions near the gelation threshold~\cite{trappe2001jamming,schenker2012influence}. We note that the variability across multiple realizations is negligible (note that the error bars are smaller than the marker size in Fig.~\ref{fig:mono}b), which emphasizes the general robustness of the simulation results. Overall, these observations confirm that, given $\ljstiffness \propto \radius$ (see Eq.~\ref{eq:ktoR}), \(\shearmodulus\) is independent of $\radius$ at packing fractions well above the gelation threshold.

This observation can be understood through dimensional analysis. The shear modulus \(\shearmodulus\) arises from the superposition of pairwise interactions, each characterized by a stiffness $\ljstiffness$. To ensure dimensional consistency, \(\ljstiffness\) must be normalized by a characteristic length scale.
In a monodisperse spherical packing with an infinitesimal interaction range, the only relevant length scale is the particle radius $\radius$, leading to the proportionality
\begin{equation}
    \shearmodulus \propto \frac{\ljstiffness}{\radius} ~.
    \label{eq:dim_analysis}
\end{equation}
Since we set $\ljstiffness \propto \radius$ (see Eq.~\ref{eq:ktoR}), the dependence on $\radius$ cancels out in Eq.~\ref{eq:dim_analysis}, making the shear modulus independent of particle size. Physically, increasing \(\radius\) reduces the number of interactions per unit volume, but this effect is exactly counterbalanced by the increased effective stiffness of each pairwise interaction. As a result, the overall shear modulus remains unchanged, which is in good agreement with our simulation results.

In contrast to variations in $\radius$, the shear modulus does vary with packing fraction $\packingfraction$ (see Fig.~\ref{fig:mono}b). More specifically, denser systems have higher $\shearmodulus$ whereby we observe a linear dependence of $\shearmodulus$ of $\packingfraction$ for $\packingfraction > 0.4$ (see Fig.~\ref{fig:mono}c). The linear dependence breaks down for systems near the gelation threshold for reasons discussed above. Interestingly, data for different particle sizes collapse onto a single curve, emphasizing that the shear modulus is primarily governed by the packing fraction.

The dominant role of the packing fraction in determining the shear modulus can be rationalized using the Cauchy-Born theory for amorphous solids~\cite{born1940stability, born1954dynamical,zaccone2009elasticity}. In systems with short-range attractive forces, particles remain close to the equilibrium position of their interaction potential, allowing the total elastic energy to be approximated as a sum of pairwise contributions in the linear regime of small strains. Assuming affine deformation and considering only the longitudinal component of central forces, the shear modulus can be expressed as
\begin{equation}
     \shearmodulus \approx \sum\limits_{1 \le i < j \le \particlenumber} \frac{\ljstiffness_{i,j}(\radius_i+\radius_j)^2}{15 \boxvolume} ~,
     \label{eq:CB_general}
\end{equation}
where $\ljstiffness_{i,j}$ is the interaction stiffness between particles $i$ and $j$, $\radius_i$ and $\radius_j$ are their respective radii, and the factor $1/15$ accounts for the spherical geometry and random orientations of the interacting pairs~\cite{shaebani2012influence}. For monodisperse systems, Eq.~\ref{eq:CB_general} simplifies to 
\begin{equation}
    \shearmodulus \approx \frac{\coordinationnumber \,\packingfraction \,\ljstiffness}{10\pi\,\radius} ~,
    \label{eq:CB_final}
\end{equation}
where the number of equal interactions is approximated by $\particlenumber \coordinationnumber / 2$ with \(\coordinationnumber\) being the average coordination number.

Using Eq.~\ref{eq:ktoR}, Eq.~\ref{eq:CB_final} simplifies to $\shearmodulus \propto \coordinationnumber \packingfraction$, consistent with our numerical observations (see inset in Fig.~\ref{fig:mono}c). This supports the applicability of the Cauchy-Born rule to this system. Note that $\coordinationnumber$ is known to vary with the packing fraction~\cite{bernal1960coordination,torquato2000random,ohern2003jamming,aste2006local}, and hence the $\shearmodulus$ vs. $\packingfraction$ relationship is nonlinear. 

We determine the exact dependence of the coordination number $\coordinationnumber$ on $\packingfraction$ for our system directly from the simulation results (see  Fig.~\ref{fig:mono}d). We observe that increases in $\packingfraction$ result in larger $\coordinationnumber$. Furthermore, for $\packingfraction > 0.4$, the relation between $\packingfraction$ and $\coordinationnumber$ is unique and independent of $\radius$, which we describe by a phenomenological fit (see black dashed line in Fig.~\ref{fig:mono}d) using:
\begin{equation}
    \approxz (\packingfraction) = \frac{1}{a \packingfraction +b} ~,
    \label{eq:fit}
\end{equation}
where $a = -0.265$ and $b = 0.277$ are fitting parameters found through least-squares. This relation will be used to analyze bidisperse systems in Sec.~\ref{sec:bi}.

Finally, we note that the average coordination number $\coordinationnumber$ drops below 6 for $\packingfraction \lesssim 0.45$ (see Fig.~\ref{fig:mono}d), which indicates sub-isostatic conditions. However, it has been shown that attractive systems can still form force-bearing structures in this regime~\cite{liu2010jamming}. Hence, the system retains a finite shear modulus but deviates from the trend observed at higher $\packingfraction$ (as described by Eq.~\ref{eq:fit}). This suggests that the systems with $\packingfraction < 0.45$ approach the gelation threshold, explaining the deviations discussed above.

In summary, our numerical results demonstrated that the shear modulus of monodisperse systems of spherical particles is independent of the particle radius and primarily governed by the packing fraction and, consequently, the average coordination number. These observations provide a basis for our investigation of bidisperse systems, which we undertake in the following section.

\section{Bidisperse particle system} \label{sec:bi}

Building on the monodisperse analysis -- where packing fraction and coordination number emerged as the primary factors governing the shear modulus -- we now extend our study to bidisperse systems, which consist of particles of two different sizes. Previous studies~\cite{mascioli_defect_2017,kumar_tuning_2016} have shown that bidisperse systems exhibit an increased elastic modulus. This effect is typically attributed to a filling mechanism, in which smaller particles occupy the voids between larger ones, allowing the system to reach higher packing fractions than its monodisperse counterpart. However, introducing additional particles simultaneously alters both the packing fraction and the microstructure, making it difficult to isolate the individual contributions of filling and structural effects. Here, we take a systematic approach to disentangle these effects by maintaining a constant packing fraction across different mixtures and analyzing the resulting elastic properties.

\subsection{Problem statement \& method} \label{sec:bi:problem_statement}

The bidisperse system consists of spherical particles  of two distinct sizes, with $\particlenumberL$ large particles of radius $\radiusL$ and $\particlenumberS$ small particles of radius $\radiusS$. The total packing fraction is given by $\packingfraction = (\volumeL + \volumeS) / \boxvolume$, where $\volumeL = 4 \pi \particlenumberL \radiusL^3 / 3$ and $\volumeS = 4 \pi \particlenumberS \radiusS^3 / 3$ are the total volume of the large and small particles, respectively. 

To account for the size disparity in particle interactions, we define the interaction radius for a particle pair $i$ and $j$ as $\radius = (\radius_i + \radius_j) / 2$. The interaction potential remains the same as in the monodisperse systems (see Eq.~\ref{eq:LJ}), but the  energy well depth now scales with the harmonic mean of the pairwise radii: $\ljeps=\scalingparameter^2\times2 (\radius_i \radius_j) / (\radius_i + \radius_j)$. This results in a stiffness scaling of $\ljstiffness\propto 2 (\radius_i \radius_j) / (\radius_i + \radius_j)$. This expression reduces to the monodisperse relation given by Eq.~\ref{eq:ktoR} when $\radius_i = \radius_j$, ensuring consistency between the two systems. 

In our simulations, we vary both the relative volume fraction of small particles, defined by
\begin{equation}
    \smallparticlevolumefraction=\frac{\volumeS}{\volumeS + \volumeL} ~,
    \label{eq:small_particle_v_fraction}
\end{equation}
and the particle size ratio, defined by
\begin{equation}
    \normalizedsmallradius=\frac{\radiusL}{\radiusS} ~.
    \label{eq:particle_size_ratio}
\end{equation}
With these two non-dimensional quantities, we can systematically explore the effect of bidispersity on the microstructure and the emergent macroscopic shear modulus.

\subsection{Results: elasticity of bidisperse particle systems} \label{sec:bi:results}

\begin{figure}
    \centering
    \includegraphics[width=1\linewidth]{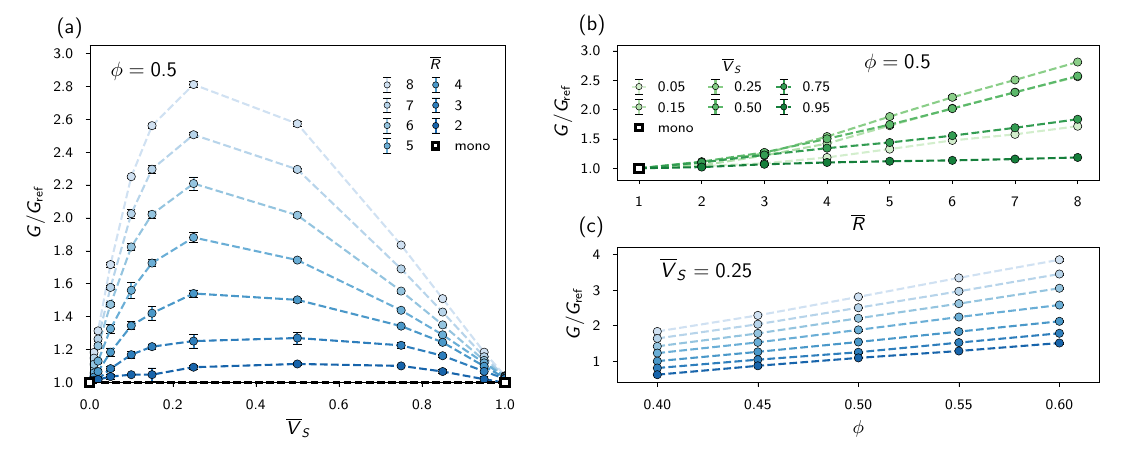}
    \caption{The shear modulus of bidisperse particle systems exceeds that of monodisperse systems at the same packing fraction and is influenced by both the particle size ratio and the relative volume fraction of small particles. \textbf{(a)}~Shear modulus \(\shearmodulus\), normalized by the reference shear modulus \(\shearmodulus_\mathrm{ref}\) (calculated as the average shear modulus of monodisperse systems at packing fraction $\packingfraction = 0.5$ and radius $\radius/\boxsize = 0.005$), plotted as a function of the relative volume fraction of small particles \(\smallparticlevolumefraction\) for various particle size ratios \(\normalizedsmallradius\) at $\packingfraction = 0.5$. Reported values are the average over 10 realizations, and error bars represent the 95\% confidence interval. \textbf{(b)} Normalized shear modulus plotted as a function of $\normalizedsmallradius$ at $\packingfraction = 0.5$ for various $\smallparticlevolumefraction$. \textbf{(c)}~Normalized shear modulus as a function of packing fraction $\packingfraction$ at $\smallparticlevolumefraction=0.25$ for different $\normalizedsmallradius$ (see legend in panel a). }
    \label{fig:bi_macro}
\end{figure}

We consider a bidisperse system with a fixed packing fraction of $\packingfraction = 0.5$ and vary the relative volume fraction of small particles, $\smallparticlevolumefraction$, across different size ratios $\normalizedsmallradius$. As large particles are gradually replaced by small ones (\ie, increasing $\smallparticlevolumefraction$), the shear modulus systematically increases for relatively small $\smallparticlevolumefraction$ (see Fig.~\ref{fig:bi_macro}a). For each $\normalizedsmallradius$, the shear modulus reaches a maximum at a certain $\smallparticlevolumefraction$, beyond which further increases in $\smallparticlevolumefraction$ lead to a decline in $\shearmodulus$, eventually converging to the monodisperse reference value for a system composed entirely of small particles ($\smallparticlevolumefraction = 1$). Consequently, at a given packing fraction, the shear modulus of bidisperse systems exceeds that of monodisperse systems for \emph{all} values of $\smallparticlevolumefraction$ and $\normalizedsmallradius$. 

The magnitude of this shear modulus increase depends on the particle size ratio. Specifically, the effect is more pronounced for systems with larger $\normalizedsmallradius$ (see Fig.~\ref{fig:bi_macro}b). This suggests that size disparity plays a crucial role in enhancing elasticity in bidisperse systems.

Beyond size ratio effects, we also examine the influence of packing fraction $\packingfraction$. For a given $\smallparticlevolumefraction$, the shear modulus increases nearly linearly with $\packingfraction$, with the rate of increase being more pronounced for larger $\normalizedsmallradius$ (see Fig.~\ref{fig:bi_macro}c). In monodisperse systems, packing fraction is the primary factor governing elasticity, and our results show, as expected, that this primary dependence persists in bidisperse systems, albeit with additional contributions from particle size disparity.

In summary, our numerical results show that the shear modulus in bidisperse systems of spherical particles is governed not only by the packing fraction $\packingfraction$ but also by the particle size ratio $\normalizedsmallradius$ and the relative volume fraction of small particles $\smallparticlevolumefraction$. However, the microstructural features responsible for the observed increase in $\shearmodulus$ remain unknown.

\subsection{Stiffening mechanisms in bidisperse particle systems} \label{sec:bi:mechanism}

To provide a mechanistic explanation of the observed increase of shear modulus in bidisperse particle systems, we now present simple models that describe the microstructure emerging from the mixing of large and small particles. Specifically, we consider two regimes, each asymptotically close to a monodisperse system. 

\begin{figure}
    \centering
    \includegraphics[width=0.85\linewidth]{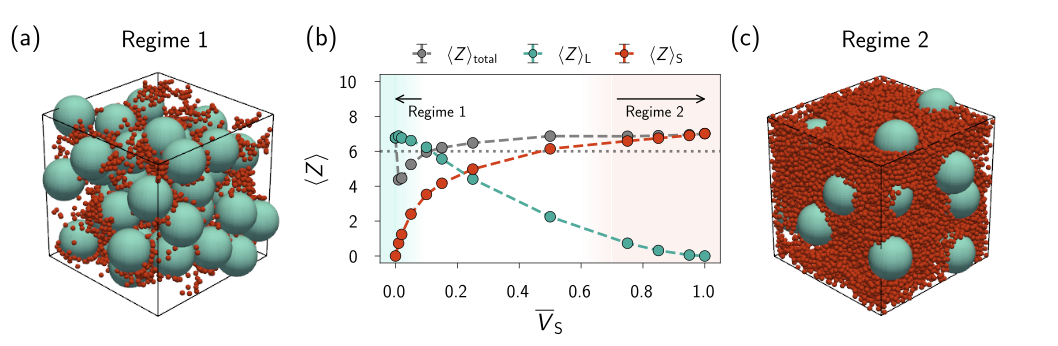}
    \caption{At extreme values of volume fraction of small particles, $\smallparticlevolumefraction \approx 0$ and $\smallparticlevolumefraction \approx 1$, the microstructure consists of a monodisperse network perturbed by the complementary particle population. \textbf{(a)}~3D visualization of an equilibrated bidisperse system at a relatively small volume fraction of small particles $\smallparticlevolumefraction=0.01$ (Regime 1). The packing fraction is $\packingfraction = 0.5$ and the particle size ratio $\normalizedsmallradius=8$. \textbf{(b)}~Average coordination numbers of total $\coordinationnumber_\mathrm{total}$, large-to-large $\coordinationnumber_\mathrm{L}$ and small-to-small $\coordinationnumber_\mathrm{S}$ particle pairs as a function of $\smallparticlevolumefraction$ at $\packingfraction=0.5$ with $\normalizedsmallradius=8$. Reported values are the averages over 10 realizations, and error bars represent the 95\% confidence interval. The dotted horizontal line corresponds to isostatic conditions. \textbf{(c)}~3D visualization of an equilibrated bidisperse system at a relatively large volume fraction of small particles $\smallparticlevolumefraction=0.75$ (Regime 2) with $\packingfraction = 0.5$ and $\normalizedsmallradius=8$.
    }
    \label{fig:bi_micro}
\end{figure}

Regime 1 corresponds to a system dominated by large particles ($\smallparticlevolumefraction \approx 0$), as illustrated in Fig.~\ref{fig:bi_micro}a. In this case, small particles remain largely dispersed without forming a continuous network, as indicated by their low coordination number $\coordinationnumber_\mathrm{S}$ (see Fig.~\ref{fig:bi_micro}b), while the primary network is composed of large particles. 

Regime 2 represents the opposite scenario, where small particles form the primary network ($\smallparticlevolumefraction \approx 1$), as shown in Fig.~\ref{fig:bi_micro}c). Here, large particles are mostly disconnected from each other, as evidenced by their low coordination number $\coordinationnumber_\mathrm{L}$ (see Fig.~\ref{fig:bi_micro}b), and do not contribute significantly to the structural connectivity. 

This framework allows us to treat each case as a perturbed monodisperse network, systematically accounting for the effects of the complementary particle population.

\subsubsection{Regime 1: large particle network at low $\smallparticlevolumefraction$} \label{sec:regime_1}

\begin{figure}
    \centering
    \includegraphics[width=0.6\linewidth]{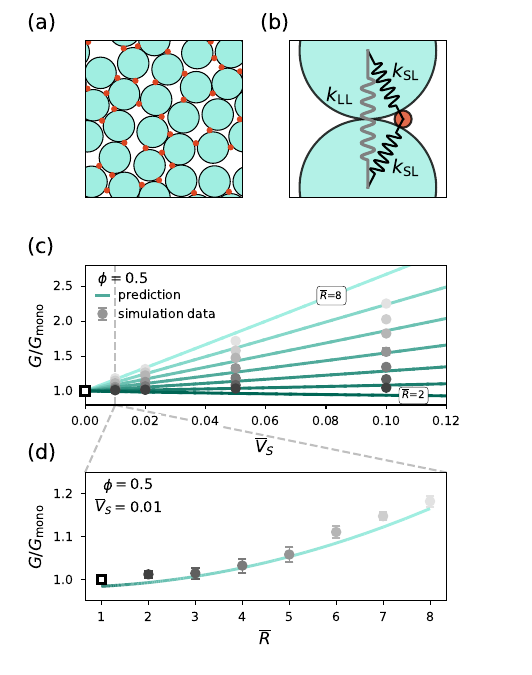}
    \caption{Effect of small particles on the large-particle network in regime 1 with $\smallparticlevolumefraction \ll 1$. The presence of small particles locally stiffens the percolated network of large particles, leading to an overall increase in shear modulus. \textbf{(a)}~Schematic 2D illustration of a bidisperse system in regime 1, showing a percolated network of large particles and small particles positioned at their junctions. \textbf{(b)}~Conceptual model illustrating how the interaction between two large particles (with stiffness $\ljstiffnessLL$) is stiffened by a small particle, effectively modeled as two additional springs of stiffness $\ljstiffnessSL$. \textbf{(c)}~Normalized shear modulus \(\shearmodulus\), relative to the average shear modulus of monodisperse systems with $\radius/\boxsize = 0.005$, plotted as a function of the relative volume fraction of small particles \(\smallparticlevolumefraction\) for various particle size ratios \(\normalizedsmallradius\) at fixed packing fraction $\packingfraction = 0.5$. Grey points represent the average values over 10 realizations (same as in Fig.~\ref{fig:bi_macro}a), and solid lines are predictions from Eq.~\ref{eq:regime1:normalizedG:simplified}, with lighter curves corresponding to larger size ratios. \textbf{(d)}~Normalized shear modulus plotted as a function of $\normalizedsmallradius$ at $\packingfraction = 0.5$ for a fixed $\smallparticlevolumefraction=0.01$.}
    \label{fig:predictive_rg1}
\end{figure}

We consider systems where the relative volume fraction of small particles approaches zero, \ie, $\smallparticlevolumefraction \approx 0$. In this regime, the large particles form a percolated network, while the small particles remain insufficient to establish their own connected structure, as illustrated in Fig.~\ref{fig:predictive_rg1}a. To explain the elastic properties of these systems, we introduce a model based on two key assumptions. First, the network of large particles serves as the primary load-bearing structure. Second, we assume that the size disparity is large enough for small particles to position themselves around the contact points of large particle pairs, stiffening their connection, as illustrated in Fig.~\ref{fig:predictive_rg1}b. We now proceed to develop this model step by step.

We begin by establishing the reference case: a monodisperse system consisting solely of large particles, \ie, $\smallparticlevolumefraction = 0$. The corresponding shear modulus can be computed using Eq.~\ref{eq:CB_final}:
\begin{equation}
    \shearmodulus\labelmono \approx \frac{\approxz(\packingfraction)  \,\packingfraction \, \ljstiffnessLL}{10\pi \radiusL} ~,
    \label{eq:GmonoL}
\end{equation}
where $\coordinationnumber$ is approximated by $\approxz(\packingfraction)$ according to Eq.~\ref{eq:fit}, and $\ljstiffnessLL$ is the interaction stiffness between two large particles at equilibrium, as given by Eq.~\ref{eq:ktoR}.

Next, we consider a bidisperse system close to this monodisperse reference case by assuming $\smallparticlevolumefraction~\ll~1$. In this regime, the large particles still form a percolated network, and we model its shear modulus using Eq.~\ref{eq:GmonoL}, with a few modifications. 

First, the large-particle network experiences a reduced packing fraction, which we account for as $\packingfractionL = \packingfraction(1-\smallparticlevolumefraction)$. Second, its coordination number follows Eq.~\ref{eq:fit}, evaluated at the effective packing fraction $\packingfractionL$. Lastly, the presence of small particles modifies the interaction stiffness between two large particles at equilibrium, an effect we denote as $\ljstiffnessLL^\mathrm{eff}$ and will determine subsequently. Incorporating these modifications, the shear modulus of the bidisperse system is expressed as
\begin{equation}
    \shearmodulus \approx \frac{\approxz(\packingfractionL)  \,\packingfractionL \, \ljstiffnessLL^\mathrm{eff}}{10\pi \radiusL} ~.
    \label{eq:regime1:basis}
\end{equation}
This approach neglects the fact that small particles occupy space that would otherwise be available to large particles, and considers their contribution solely through the stiffening of the large-particle network.

We now determine the effective stiffness of junctions in the large-particle network, denoted by $\ljstiffnessLL^\mathrm{eff}$. We assume that small particles are sufficiently small to position themselves at the junctions of (exactly) two large particles, as illustrated in Fig.~\ref{fig:predictive_rg1}b. By occupying the equilibrium position, a small particle stiffens the interaction between the two large particles -- which alone would have stiffness $\ljstiffnessLL$ -- through two additional small-large interactions of stiffness $\ljstiffnessSL$. On average, this results in an effective junction stiffness of
\begin{equation}
    \ljstiffnessLL^\mathrm{eff} = \ljstiffnessLL + \frac{\particlenumberS}{\connectionnumberLL } \frac{\ljstiffnessSL}{2} ~,
    \label{eq:regime1:kLLeff}
\end{equation}
where $\ljstiffnessSL/2$ accounts for the two in-series small-large interactions, and $\particlenumberS/\connectionnumberLL$ represents the average number of small particles per large-large particle junction in the system. This ratio can be expressed as
\begin{equation}
    \frac{\particlenumberS}{\connectionnumberLL} = \frac{2}{\approxz(\packingfractionL) } \frac{\smallparticlevolumefraction}{1-\smallparticlevolumefraction} \normalizedsmallradius^{3}  ~,
    \label{eq:regime1:NSNLL}
\end{equation}
where we use $\connectionnumberLL = \particlenumberL \approxz(\packingfractionL) / 2$ for the number of large-large junctions and express $\particlenumberS$ and $\particlenumberL$ in terms of particle radii and their respective volume fractions. 

Finally, we substitute Eqs~\ref{eq:regime1:kLLeff} and \ref{eq:regime1:NSNLL} into Eq.~\ref{eq:regime1:basis} and normalize by the monodisperse reference value given in Eq.~\ref{eq:GmonoL}, yielding
\begin{equation}
    \frac{\shearmodulus}{\shearmodulus_\mathrm{mono}} \approx(1-\smallparticlevolumefraction) \frac{\approxz(\packingfractionL) }{\approxz(\packingfraction)}+\smallparticlevolumefraction\frac{\ljstiffnessSL}{\ljstiffnessLL}\frac{  \normalizedsmallradius^{3}}{\approxz(\packingfraction)} ~.
    \label{eq:regime1:normalizedG}
\end{equation}
Assuming a large size disparity, we apply the asymptotic limit $\lim_{ \normalizedsmallradius \to \infty} \ljstiffnessSL / \ljstiffnessLL = 2 /\normalizedsmallradius$, which simplifies Eq.~\ref{eq:regime1:normalizedG} to
\begin{equation}
        \frac{\shearmodulus}{\shearmodulus_\mathrm{mono}} \approx(1-\smallparticlevolumefraction) \frac{\approxz(\packingfractionL) }{\approxz(\packingfraction)}+ \smallparticlevolumefraction \frac{2 \normalizedsmallradius^{2}  }{\approxz(\packingfraction)} ~.
        \label{eq:regime1:normalizedG:simplified}
\end{equation}
This expression provides a closed-form prediction for the shear modulus of bidisperse systems in regime 1, where a small number of small particles stiffens a network of larger ones.

We now analyze this prediction and compare it to our numerical results presented in Sec.~\ref{sec:bi:results}. For size ratios $\normalizedsmallradius \gtrsim 3$, which are consistent with the assumptions underlying the model, the prediction captures the key trend: the shear modulus increases relative to the monodisperse case for any value of $\smallparticlevolumefraction$ (see Fig.~\ref{fig:predictive_rg1}c). Moreover, the model correctly reflects the observation that larger $\normalizedsmallradius$ lead to a greater increase in shear modulus, as shown in the numerical results (see Fig.~\ref{fig:predictive_rg1}d). While the overall trends are consistent, some quantitative discrepancies arise. In particular, for small $\normalizedsmallradius$ and somewhat larger $\smallparticlevolumefraction$, the model predicts a decrease in shear modulus compared to the monodisperse case, whereas the simulations show an increase (see right region in Fig.~\ref{fig:predictive_rg1}c). Furthermore, for large $\normalizedsmallradius$, the model overestimates the increase in the shear modulus. These discrepancies and their possible origins will be discussed in Sec.~\ref{sec:discussion}.

\subsubsection{Regime 2: small particle network at high $\smallparticlevolumefraction$}

\begin{figure}
    \centering
    \includegraphics[width=0.6\linewidth]{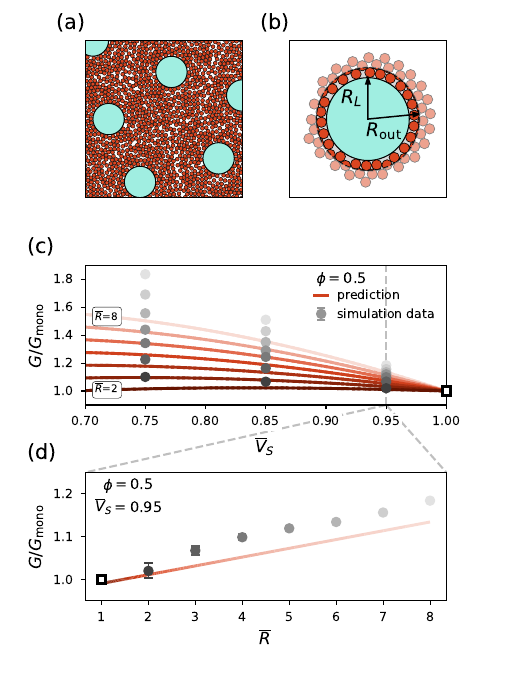}
    \caption{Effect of large particles on the small-particle network in regime 2 with $\smallparticlevolumefraction \approx 1$. The presence of large particles creates a cluster of stiff connections between many small particles and a single large particle. \textbf{(a)}~Schematic 2D illustration of a bidisperse system in regime 2, showing a percolated network of small particles with few large particles embedded. \textbf{(b)}~Conceptual model illustrating the arrangement of small particles around a large particle creating many stiff connections. A shell of inner radius $\radiusL$ and outer radius $\radius_\mathrm{out}=\radiusL+2\radiusS$ defines the area of small particles that can connect to the large particle. \textbf{(c)}~Normalized shear modulus \(\shearmodulus\), relative  to the average shear modulus of monodisperse systems with $\radius/\boxsize = 0.005$, plotted as a function of the relative volume fraction of small particles \(\smallparticlevolumefraction\) for various particle size ratios \(\normalizedsmallradius\) at fixed packing fraction $\packingfraction = 0.5$. Grey points represent the average values over 10 realizations (same as in Fig.~\ref{fig:bi_macro}a), and solid lines are predictions from Eq.~\ref{eq:regime2:normalizedG:simplified}, with darker colors corresponding to smaller size ratios. \textbf{(d)}~Normalized shear modulus plotted as a function of $\normalizedsmallradius$ at $\packingfraction = 0.5$ for a fixed $\smallparticlevolumefraction=0.95$.}
    \label{fig:predictive_rg2}
\end{figure}

We now turn to systems where the relative volume fraction of small particles is close to one, \ie, $\smallparticlevolumefraction \approx 1$. In this limit, the small particles form a percolated network, while the large particles are dispersed within it without forming a connected structure of their own, as illustrated in Fig.~\ref{fig:predictive_rg2}a. To capture the elastic response of such systems, we construct a model based on two central assumptions. First, the load-bearing structure is primarily composed of small particles. Second, we assume that the size disparity is sufficiently large for each large particle to be fully embedded within the surrounding network of small particles, as illustrated in Fig.~\ref{fig:predictive_rg2}b. In the following, we develop this model in a step-by-step manner.

As in regime 1, we begin by defining a monodisperse reference case. However, in this case, the system consists solely of small particles, \ie, $\smallparticlevolumefraction = 1$, in contrast to the large-particle network considered previously. The corresponding shear modulus is given by Eq.~\ref{eq:CB_final}:
\begin{equation}
    \shearmodulus\labelmono \approx \frac{\approxz(\packingfraction)  \,\packingfraction \, \ljstiffnessSS}{10\pi \radiusS} ~,
    \label{eq:regime2:Gmono}
\end{equation}
where $\coordinationnumber$ is approximated by $\approxz(\packingfraction)$ according to Eq.~\ref{eq:fit}, and $\ljstiffnessSS$ is the equilibrium interaction stiffness between two small particles, as given by Eq.~\ref{eq:ktoR}.

Next, we consider a bidisperse system close to this monodisperse reference case by assuming $\smallparticlevolumefraction \approx 1$. In this regime, the small particles still form a percolated network, and we describe its shear modulus based on Eq.~\ref{eq:regime2:Gmono}, with modifications to account for the presence of large particles:
\begin{equation}
    \shearmodulus \approx \frac{\approxz(\packingfractionS)  \,\packingfractionS \, \ljstiffnessSS}{10\pi \radiusS} + \shearmodulus\labelL ~,
    \label{eq:regime2:basis}
\end{equation}
where $\packingfractionS = \packingfraction \smallparticlevolumefraction$ is the nominal packing fraction of small particles, calculated without accounting for the volume excluded by the large particles. The term $\shearmodulus\labelL$ captures the additional stiffness contribution from the embedded large particles and will be determined in the following. 

Since only a small number of large particles are introduced in this regime, we assume that their primary contribution arises from forming additional small-large interactions. These interactions can be estimated using Eq.~\ref{eq:CB_general} as
\begin{equation}
    \shearmodulus\labelL = \frac{1}{15} \frac{\connectionnumberSL}{\boxvolume} \ljstiffnessSL \left( \radiusS + \radiusL \right)^2 ~,
    \label{eq:regime2:GL}
\end{equation}
where $\connectionnumberSL$ is the total number of small-large particle interactions. Next, we estimate $\connectionnumberSL$, using the following assumptions: 1) only small particles located within a distance of $\radius_\mathrm{out} = \radiusL + 2\radiusS$ from the center of a large particle can establish such a connection (see Fig.~\ref{fig:predictive_rg2}b), 2) each small particle within this interaction range forms exactly one connection with the neighboring large particle, and 3) the local packing fraction of small particles is unaffected by the presence of large particles. Under these assumptions, the total number of small-large interactions in the system is given by: 
\begin{equation}
    \connectionnumberSL = \frac{\volume_\mathrm{shell}}{\boxvolume} \particlenumberS ~,
    \label{eq:regime2:NSL}
\end{equation}
where $\volume_\mathrm{shell} = 4\pi \particlenumberL (\radius_\mathrm{out}^3 - \radiusL^3) / 3$ is the total volume of the ``interaction'' shells around \emph{all} large particles in the system (see Fig.~\ref{fig:predictive_rg2}b). 

Finally, we substitute Eqs.~\ref{eq:regime2:GL} and \ref{eq:regime2:NSL} into Eq.~\ref{eq:regime2:basis} and normalize by the monodisperse reference value given in Eq.~\ref{eq:regime2:Gmono}, resulting in
\begin{equation}
    \frac{\shearmodulus}{\shearmodulus\labelmono} = \smallparticlevolumefraction \frac{\approxz(\packingfractionS)}{\approxz(\packingfraction)} + (1 - \smallparticlevolumefraction) \frac{\packingfractionS }{\approxz(\packingfraction)} f(\normalizedsmallradius) ~,
    \label{eq:regime2:normalizedG}
\end{equation}
where $f(\normalizedsmallradius) = 6 \normalizedsmallradius + 18 + 20\normalizedsmallradius^{-1} + 8\normalizedsmallradius^{-2}$. Assuming a large size disparity, we simplify $f(\normalizedsmallradius)$ to its leading-order terms, which simplifies Eq.~\ref{eq:regime2:normalizedG} to
\begin{equation}
    \frac{\shearmodulus}{\shearmodulus\labelmono} = \smallparticlevolumefraction \frac{\approxz(\packingfractionS)}{\approxz(\packingfraction)} + (1 - \smallparticlevolumefraction) \frac{ (6\normalizedsmallradius + 18) }{\approxz(\packingfraction)} \packingfractionS ~.
    \label{eq:regime2:normalizedG:simplified}
\end{equation}
This provides a closed-form prediction for the shear modulus in bidisperse systems in regime 2, where a small number of large particles stiffen a percolated network of small ones.

We now compare this prediction with the numerical results presented in Sec.~\ref{sec:bi:results}. The model captures the key trends observed in the simulations: the shear modulus increases relative to the monodisperse case as $\smallparticlevolumefraction$ decreases (see Fig.~\ref{fig:predictive_rg2}c), and the magnitude of this increase is greater for larger values of $\normalizedsmallradius$ (see Fig.~\ref{fig:predictive_rg2}d). While the overall trends are consistent between prediction and simulation, some quantitative discrepancies remain. In particular, the prediction underestimates the shear modulus for systems with relatively small $\normalizedsmallradius$ (see left region of Fig.~\ref{fig:predictive_rg2}c). Possible reasons for these discrepancies will be discussed in Sec.~\ref{sec:discussion}.

\section{Discussion} \label{sec:discussion}

Our study demonstrates that introducing bidispersity into particle systems with attractive interactions leads to a systematic increase in shear modulus compared to the monodisperse case, even when the total packing fraction is held constant. Numerical simulations reveal that this stiffening effect depends on both the particle size ratio and the relative volume fraction of small particles. 
To explain these trends, we developed two asymptotic models corresponding to the limiting cases of low and high content of small particles, referred to as regime 1 and 2, respectively. In regime 1, a percolated network of large particles is locally stiffened by the presence of a few small particles. In regime 2, a few large particles are embedded within a percolated network of small particles. For both regimes, we derived closed-form predictions that capture the qualitative behavior observed in simulations, including the dependence of the shear modulus on particle size ratio and composition.

The observed increase in shear modulus arises from microstructural changes introduced by mixing particles of different sizes. In regime 1, small particles position themselves at the junctions between large particles, effectively stiffening large–large contacts. In regime 2, large particles act as inclusions that increase local constraints within the small-particle network. Despite the low volume fraction of the minority species in both cases, their geometric placement and interaction stiffness are sufficient to alter the mechanical response of the entire system. These findings suggest that bidispersity can be used not only to tune packing efficiency but also to control elasticity through localized structural stiffening.

To derive closed-form predictions, models for both regimes rely on simplifying assumptions that enable analytical tractability but may limit quantitative accuracy. One such assumption is that the packing fraction of the majority species is computed without accounting for the volume occupied by the minority species. This simplification neglects excluded volume effects and may lead to underestimating the local density of the load-bearing network. A more accurate approach would define an effective packing fraction that considers only the free volume available to the majority species. For instance, in regime 2, the effective packing fraction of the small particles can be written as
\begin{equation} 
    \packingfractionS^* = \frac{\volumeS}{\boxvolume - \volumeL} ~,
\end{equation}
where $\volumeS$ is the total volume of small particles and $\volumeL$ is the total volume of large particles. Using such corrected values could improve the accuracy of the predicted shear modulus, especially at higher volume fractions of the minority species, but at the cost of clarity of the final closed-form expression. 

Another important limitation shared by both models is their inability to describe systems with intermediate mixing ratios, where the volumes of small and large particles are comparable. In this regime, neither species occupies sufficient volume to form a percolated network, violating the core assumption underlying both models. As a result, the mechanical response cannot be captured by either asymptotic framework. Addressing this regime would require alternative modeling approaches that explicitly account for the mixed connectivity of the system. One possible route is to perform a network-based analysis of the connected structure, for example by identifying the percolating backbone using cluster analysis or computing species-resolved connectivity graphs. Quantities such as the largest connected component, average path length, or degree distribution for each particle type could offer insight into which substructures contribute to stiffness and how load-bearing pathways emerge in the absence of a dominant phase~\cite{newman2018networks,schenker2012influence,majmudar2005contact}.

In regime 1, the model assumes a large size disparity, allowing small particles to position themselves at the junctions between large particles and stiffen the network. However, when the size ratio is too close to one, this arrangement is no longer geometrically feasible, and the small particles cannot occupy the idealized equilibrium positions. This results in a different microstructure, where the stiffening mechanism assumed by the model does not apply. Moreover, the model assumes that all small particles contribute to reinforcing distinct large–large contacts. At higher small-particle volume fractions, however, the number of available positions at junctions becomes limited. Once these are saturated, excessive small particles must occupy other regions in the structure, often clustering locally within voids instead of forming well-distributed interstitial connections, and hence establish softer links. As a result, the model tends to slightly overestimate the shear modulus, particularly for smaller size ratios and larger small-particle fractions. A more refined model could include saturation effects and explicitly account for the spatial distribution and connectivity of small particles beyond idealized pairwise reinforcement.

In regime 2, the model assumes that a small number of large particles are embedded in a percolated network of small particles, contributing to stiffness through additional small–large interactions. While the model captures the qualitative trends, it systematically underestimates the shear modulus. We attribute this overall discrepancy to the underestimation of effective packing fraction, as discussed above. Additional discrepancies at small particle size ratios are expected due to the asymptotic approximation used for the function $f(\normalizedsmallradius)$. In addition, the model accounts only for the small–large interactions introduced by the presence of large particles but neglects their effect on the connectivity of the small-particle network. In practice, large particles influence the spatial arrangement of surrounding small particles, potentially decreasing the number of small–small contacts and weakening the overall stiffness. This effect is not captured in the current formulation and could be incorporated in future models through a corrective term that accounts for changes in small-particle coordination in the vicinity of large inclusions.

Finally, while the models were developed for idealized bidisperse systems with attractive interactions, the underlying mechanisms of microstructural stiffening are broadly applicable. They suggest that bidispersity offers a robust strategy for tuning the elastic properties of particle-based materials beyond what is achievable through packing fraction alone. This has potential implications for a range of soft materials, including colloidal suspensions, where controlling stiffness through compositional or geometric design is of practical importance.

\section{Conclusion} \label{sec:conclusion}

We have investigated how bidispersity influences the elastic properties of particle systems with short-range attractive interactions. Using numerical simulations at fixed total packing fraction, we showed that introducing a second particle size systematically increases the shear modulus relative to monodisperse systems. The extent of stiffening depends on both the particle size ratio and the volume fraction of the minority species. To explain these trends, we developed two asymptotic models corresponding to limiting cases in which either the large or the small particles form a percolated network. Each model provides a closed-form prediction for the shear modulus based on microstructural considerations and captures the key features observed in simulations. Together, these findings demonstrate that bidispersity offers a simple and effective strategy to tune the mechanical response of particle-based materials through controlled variation in size and composition.

\section{Acknowledgements}
The authors acknowledge the Swiss National Science Foundation for financial support under grant number 200021\_200343. They also thank Michal Hlobil for helpful discussions and Alessandra Lingua and Mohit Pundir for feedback on the manuscript draft.

\section{CRediT authorship contribution statement}
\textbf{Yaqi Zhao}: Conceptualization, Methodology, Validation, Investigation, Formal Analysis, Data Curation, Visualization, Writing -- Original Draft. \textbf{Antoine Sanner}: Conceptualization, Formal analysis, Supervision, Writing -- Review \& Editing. \textbf{Luca Michel}: Conceptualization, Visualization, Writing -- Review \& Editing. \textbf{David S. Kammer}: Conceptualization, Formal analysis, Supervision, Writing -- Review \& Editing, Funding acquisition.

\section{Declaration of competing interest}
The authors declare that they have no known competing financial interests or personal relationships that could have appeared to influence the work reported in this paper.

\section{Code Availability}
The code used for the numerical simulations is available on ETH GitLab: [URL]

\section{Data Availability}
The generated data has been deposited in the ETH Research Collection database under accession code [URL]

\appendix

\section{Solver Details} \label{appendix:solver}
All simulations were performed using the LAMMPS software package~\cite{LAMMPS} with the 15Jun2023 patch. The parameters for the generation of a non-overlapping initial configuration of particles through a soft relaxation process are provided in Table~\ref{tab:init_params}, and for the equilibration simulations are given in Table~\ref{tab:equilibration_params}. 

\begin{table}[htbp]
    \centering
    \caption{Summary of key parameters used for the initial placement and soft relaxation.}
    \label{tab:init_params}
    \begin{tabular}{p{0.45\linewidth} p{0.45\linewidth}}
    \hline
    \textbf{Parameter} & \textbf{Value / Setting} \\
    \hline
    Integration & \verb|nve/sphere| \\
    Boundary Condition & Periodic \\
    \hline
    \textbf{Time Stepping and Run Duration} &  \\
    \quad Time Step Size & 0.0001 \\
    \quad Number of Time Steps & 1,000,000 \\
    \hline
    \textbf{Soft Repulsive Model} & \verb|pair_style hooke| \\
    \quad Normal Stiffness & $k_n = 10,000$ \\
    \quad Normal Damping & $\eta = 0$ \\
    \quad Tangential Model & \verb|linear_history| \\
    \quad Tangential Stiffness & $k_t = 7,000$ \\
    \quad Tangential Damping & 0 \\
    \quad Tangential Friction Coefficient & 0.1 \\
    \hline
    \textbf{Damping Parameters} &  \\
    \quad Viscous Damping & \verb|viscous 1.88| \\
    \quad Scaling Factor & \verb|scale 2 size_ratio| \\
    & (Damping coefficient scaled by $\normalizedsmallradius$) \\
    \hline
    \end{tabular}
\end{table}

\begin{table}[htbp]
    \centering
    \caption{Summary of key parameters used for the equilibration stage.}
    \label{tab:equilibration_params}
    \begin{tabular}{p{0.45\linewidth} p{0.45\linewidth}}
    \hline
    \textbf{Parameter} & \textbf{Value / Setting} \\
    \hline
    Time Step Size & 0.0001 \\
    \hline
    \multicolumn{2}{l}{\textbf{Energy Minimization}} \\
    \quad Minimization Method & FIRE \\
    \quad Convergence Criteria & \verb|1e-10 1e-10 1e7 1e7| \\
    \quad \quad Energy Tolerance & $1.0 \times 10^{-10}$ \\
    \quad \quad Force Tolerance & $1.0 \times 10^{-10}$ \\
    \quad \quad Maximum Iterations & 10,000,000 \\
    \quad \quad Maximum Force Evaluations & 10,000,000 \\
    \hline
    \end{tabular}
\end{table}

\section{Shear modulus computation} \label{appendix:modulus_computation}

The shear modulus is determined by evaluating the second derivatives of the total potential energy with respect to strain, yielding the Born tensor~\cite{theodorou1986atomistic,lutsko1989generalized}:
\begin{equation}
    C^B_{i,j} = \frac{1}{\boxvolume} \frac{\partial^2U}{\partial\varepsilon_i\partial\varepsilon_j}=\frac{1}{\boxvolume} \sum_{\alpha < \beta} 
\left[
\frac{d^2 U_{\alpha\beta}}{d r_{\alpha\beta}^2} 
\frac{\partial r_{\alpha\beta}}{\partial \varepsilon_i} 
\frac{\partial r_{\alpha\beta}}{\partial \varepsilon_j} 
+ \frac{d U_{\alpha\beta}}{d r_{\alpha\beta}} 
\frac{\partial^2 r_{\alpha\beta}}{\partial \varepsilon_i \partial \varepsilon_j} 
\right]
  ~,
\end{equation}
where $\varepsilon_i$ and $\varepsilon_j$ denote the components of the strain tensor $\boldsymbol{\varepsilon}$, $r_{\alpha\beta} = \lvert \mathbf{r}_{\beta} - \mathbf{r}_{\alpha} \rvert $ is the distance between particles $\alpha$ and $\beta$, and $U_{\alpha\beta}$ is the pairwise potential energy. 

The first derivative of distance $r_{\alpha\beta}$ w.r.t. strain tensor component is given by:
\[
\frac{\partial r_{\alpha\beta}}{\partial \varepsilon_i} = \frac{\mathbf{r}_{\alpha\beta} \cdot \mathbf{n}_i}{r_{\alpha\beta}}, 
\]
and the second derivative is:
\[
\frac{\partial^2 r_{\alpha\beta}}{\partial \varepsilon_i \partial \varepsilon_j} = \frac{1}{r_{\alpha\beta}} 
\left[ 
\mathbf{n}_i \cdot \mathbf{n}_j - 
\left( \frac{\mathbf{r}_{\alpha\beta} \cdot \mathbf{n}_i}{r_{\alpha\beta}} \right) 
\left( \frac{\mathbf{r}_{\alpha\beta} \cdot \mathbf{n}_j}{r_{\alpha\beta}} \right) 
\right],
\]
where $ \mathbf{n}_i$ is the unit vector along the direction of deformation.

\end{document}